\documentclass[pre,aps,twocolumn,showpacs,amsmath,amssymb,amsfonts]{revtex4}
\usepackage{epsfig}
\usepackage{xspace}
\usepackage{amscd}
\usepackage{bm}
\newcommand{\pcpdt}{PCPD2\xspace}
\newcommand{\Cpcpdt}{PCPD-DI\xspace}
\newcommand{\dpt}{{\tt tp22}\xspace}
\newcommand{\Cdpt}{DP-DI\xspace}
\begin{document}
\title{Crossover from the parity-conserving pair contact process with diffusion
to other universality classes}
\author{Su-Chan Park}
\affiliation{Institut f\"ur Theoretische Physik, Universit\"at zu K\"oln,
Z\"ulpicher Str. 77, 50937 K\"oln, Germany}
\author{Hyunggyu Park}
\affiliation{School of Physics, Korea Institute for Advanced Study, Seoul 130-722, Korea}
\date{\today}
\begin{abstract}
The pair contact process with diffusion (PCPD) with modulo 2 conservation (\pcpdt)
[$2A\rightarrow 4A$, $2A\rightarrow 0$] is studied in one dimension,
focused on the crossover to other well established universality classes:
the directed Ising (DI) and the directed percolation (DP).
First, we show that the \pcpdt shares the critical behaviors with the PCPD, both with and
without directional bias. Second, the crossover from the \pcpdt to the DI is studied by including
a parity-conserving single-particle process ($A \rightarrow 3A$). We find
the crossover exponent $1/\phi_1 = 0.57(3)$, which is argued to be identical to that of
the PCPD-to-DP crossover by adding $A \rightarrow 2A$.  This suggests that the
PCPD universality class has a well defined fixed point distinct from the DP.
Third, we study the crossover from a hybrid-type reaction-diffusion process
belonging to the DP [$3A\rightarrow 5A$, $2A\rightarrow 0$] to the DI by adding $A \rightarrow 3A$.
We find $1/\phi_2 = 0.73(4)$ for the DP-to-DI crossover. The inequality of $\phi_1$ and $\phi_2$ further supports the non-DP nature of the PCPD scaling.
Finally, we introduce a symmetry-breaking
field in the dual spin language to study the crossover from the \pcpdt to the DP. We find
$1/\phi_3 = 1.23(10)$, which is associated with a new independent route from the PCPD to the DP.

\end{abstract}
\maketitle
\section{\label{Sec:intro}introduction}
The pair contact process with diffusion (PCPD) is an interacting particle system
with diffusion, pair annihilation ($2A\rightarrow 0$), and creation of offspring by pairs
($2A\rightarrow 3A$)~\cite{HH04,PP08a,HT97}. It is well known that the PCPD scaling is definitely distinct
from the directed percolation (DP) scaling in two dimensions and higher. However, for the one dimensional
PCPD, there is still no consensus on its critical scaling with two different viewpoints such as
the DP scaling after an extremely long crossover time~\cite{BC03,Hin06,SB08} and the new scaling distinct from the DP~\cite{PP08a,PP07,PP06,PP05a,PP05b,NP04,KC03,JvWDT04}.
Since the critical exponents of the PCPD are so similar
to those of the DP in one dimension and the strong corrections to scaling are present,
it would be difficult, if not impossible, to settle down the controversy via
direct numerical simulations of the PCPD.
On this account, the authors suggested two critical
tests in the past. One is the driven PCPD (DPCPD)~\cite{PP05a,PP05b} and the other is the
crossover~\cite{PP06,PP07}. A recent brief review on
these two approaches is available in Ref.~\cite{PP08a}.
In this paper, we continue to investigate the crossover behavior of the PCPD to other universality classes,
in order to understand further the difference between the PCPD and the DP.

It is well established that the critical behavior of the PCPD is not affected by
introducing modulo 2 (parity) conservation of the total particle number~\cite{PHK01,KC03}.
That is, even if a
pair branches two offspring ($2A\rightarrow 4A$) rather than one,
this model is believed to share the critical behavior with the PCPD and does not
belong to the directed Ising (DI)~\cite{KP94,PP95,HKPP98} or the parity-conserving (PC)
class~\cite{GKvdT84,Gra89}.
In the context of the crossover, however, the parity-conserving PCPD provides
a useful platform to study rich crossover behaviors compared to the non-conserving PCPD.
Because of the special role of the parity conservation,
we will refer to the parity-conserving PCPD as \pcpdt  in this paper.
The term PCPD will be preserved to refer to the universality class in what follows.\

We introduce a single-particle branching process in the PCPD2 to study the crossover behavior
from the PCPD to other universality classes.
Depending on the number of offspring in the single-particle process,
one can study the crossover either to the DP (by $A \rightarrow 2A$)
as in Ref.~\cite{PP06}, or to the DI (by $A\rightarrow 3A$; see Sec.~\ref{sub:PCPD-DI}).
Since the PCPD is clearly independent from the DI class, we expect a nontrivial crossover scaling
from the PCPD to the DI. However, this crossover scaling may not be independent from the PCPD-to-DP
crossover scaling, if the PCPD universality class can be described by a well-defined fixed point
and the parity conservation is irrelevant to the PCPD fixed point.

As discussed in our previous work~\cite{PP08b} and also in Sec.~\ref{sub:PCPD-DI},
the crossover scaling depends only on the initial fixed point and its crossover operator, but
not directly on the terminal fixed point. The initial PCPD fixed point is blind to the parity conservation~\cite{PHK01,KC03}. Therefore, in the PCPD point of view, the crossover operators associated with $A \rightarrow 2A$ and $A \rightarrow 3A$ may be undistinguishable, and thus have the same scaling dimension.
Hence, we expect the same crossover exponent for the PCPD-to-DP and the PCPD-to-DI crossover scaling,
which is numerically confirmed in Sec.~\ref{sub:PCPD-DI}. The terminal fixed point (DP or DI), of course, depends crucially on the existence of the parity conservation in the reaction dynamics.

More complex reaction-diffusion models with hybrid-type dynamics may be considered such that
$mA\rightarrow (m+k)A$ and $nA\rightarrow (n-l)A$. These models are numerically shown~\cite{KC03}
and reasonably argued [see Sec.~V in Ref.~\cite{PP07}] to belong to the DP in one dimension when $m>n$ and $k,l>0$, in contrast to the cases of $m=n$ (PCPD at $m=n=2$ and TCPD at
$m=n=3$~\cite{PHK02,KC03}). The crossover from the
diffusing {\tt tp12} $(3A\rightarrow 4A, 2A\rightarrow 0)$ to the DP by adding a single-particle reaction ($A\rightarrow 0$) has been studied previously~\cite{PP07} to find
the typical ``crossover'' behavior between the identical (DP) universality classes as expected: a linear phase boundary
and no diverging time scale. In this paper, we study the crossover from the diffusing {\tt tp22} $(3A\rightarrow 5A, 2A\rightarrow 0)$ with the parity conservation to the DI by adding $A\rightarrow 3A$. A nontrivial crossover exponent for this DP-to-DI crossover is numerically estimated, which is different from the PCPD-to-DI crossover exponent. This finding again supports the distinct nature of the PCPD scaling from the DP.

When $A\rightarrow 2A$ is added to {\tt tp22}, we again observe the trivial crossover between the identical (DP) classes. One may question how the {\tt tp22} can distinguish the crossover operators associated with $A\rightarrow 2A$ and $A\rightarrow 3A$, in contrast to the PCPD. The answer is simply that the parity
conservation at the level of single-particle reactions is relevant to the DP fixed point, but not to the PCPD fixed point. This again further supports
the difference between the DP and the PCPD fixed point even in one dimension.

The parity conservation in one dimension makes it possible to map the \pcpdt to
the nonequilibrium Ising spin dynamics with the $Z_2$ symmetry~\cite{KP94,HKPP98,PP08b}.
Although this mapping can be considered nominal due to the irrelevance of the parity conservation
in the PCPD, one can still ask a question regarding the effect of the symmetry breaking field in the
dual spin language. In terms of the original particle (kink) language, it implies an alternating directional
bias in diffusion (or branching) of particles, which makes a pair of neighboring particles tightly bound.
Then we can easily expect a simple DP-type dynamics of bound pairs ($X\rightarrow 2X$, $X\rightarrow 0$).
Thus, with the symmetry breaking field, we may find a new  independent route from the PCPD
to the DP, which is characterized by a new crossover exponent. (Similar independent routes
are discussed for the DI-to-DP crossover in Ref.~\cite{PP08b}). This is also confirmed by numerical simulations.

This paper is organized as follows: Section~\ref{Sec:mod2} introduces the dynamic
rules of the \pcpdt and shows that the \pcpdt shares the critical behavior with
its cousin without conservation (PCPD). By introducing the relative bias between
isolated particles and pairs, the \pcpdt exhibits the same critical scaling
as in the DPCPD. In Sec.~\ref{Sec:models},
we introduce the three crossover models by adding a single-particle reaction process
to the PCPD2 and to \dpt, and by adding an alternating directional bias to the PCPD2, respectively.
In Sec.~\ref{Sec:Numerics}, the extensive numerical results are presented for these three models. Finally, the conclusion and the summary of the results follow in
Sec.~\ref{Sec:sum}.

\section{\label{Sec:mod2}\pcpdt}

This section begins with the description of the \pcpdt dynamics
adopted for simulations, and then reconfirms the PCPD behavior of the \pcpdt.
The dynamic rules for the \pcpdt are as follows:
\begin{eqnarray}
\label{Eq:pcpd}
A\emptyset \stackrel{D}{\rightarrow} \emptyset A,
\;
\emptyset A\stackrel{D}{\rightarrow}A \emptyset ,
\;
A A \stackrel{p}{\rightarrow} \emptyset \emptyset,
\\
A A \emptyset \emptyset \stackrel{\sigma_R}{\rightarrow} AAAA, \quad
\emptyset \emptyset A A \stackrel{\sigma_L}{\rightarrow} AAAA,
\nonumber
\end{eqnarray}
where $A$ ($\emptyset$) stands for an occupied (vacant) site and the
parameters on the top of arrows represent the transition rates. For convenience, we set
$\sigma_R + \sigma_L = 1-p$ with $0 \le p \le 1$.
Except the biased branching model (see below), $\sigma_R = \sigma_L$ is set
in most cases.
Each site can carry at most one particle (hard core exclusion).

The \pcpdt has been shown numerically to belong to the PCPD class~\cite{PHK01,KC03}. This may be understood
by viewing the PCPD-type models  as a coupled system of two particle species,
where a particle pair corresponds to an $X$ particle and a solitary particle to a $Y$ particle~\cite{Hin01a,NP04,com1}.
Then, the PCPD dynamics is the combination of the DP-like dynamics of $X$'s and the binary annihilation
diffusing dynamics of $Y$'s. The two species are coupled through transmutations.
The only difference in the PCPD2 dynamics from the PCPD is the number of offspring in branching of $X$ particles such as $X\rightarrow 2X$ in the PCPD and $X\rightarrow 3X$ in the PCPD2, which should be
irrelevant in the DP dynamics of $X$'s with no parity conservation by $X\rightarrow 0$. Therefore,
it is not surprising to find the PCPD2 in the PCPD class.

In this paper, we numerically study the PCPD2 with directional bias and check whether it shares the same critical behavior with the driven PCPD (DPCPD)~\cite{PP05a,Psc06}. This may serve as another evidence supporting that the parity conservation in the PCPD2 is irrelevant.
Being relevant the relative bias between isolated
particles and pairs, the biased branching as well as the biased diffusion
should entail the change of the universality class~\cite{PP05a}.
Since the data with the biased branching have not been reported in the literature,
we consider here the biased branching for the PCPD2 by setting
$\sigma_L=0$ and $\sigma_R = 1-p$ in Eq.~\eqref{Eq:pcpd}.
The diffusion constant is set as $D=\frac{1}{2}$.

\begin{figure}[t]
\includegraphics[width=0.45\textwidth]{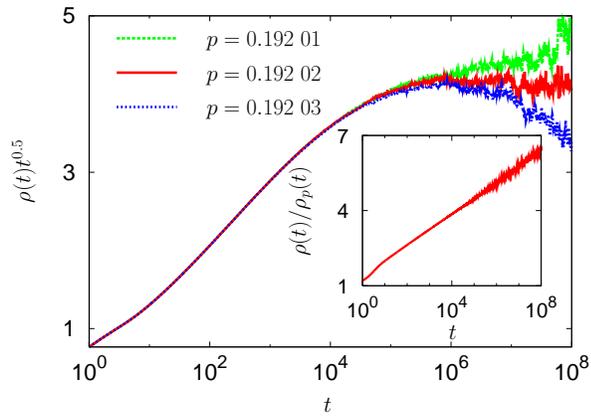}
\caption{\label{Fig:DPCPD} (Color online) Semilogarithmic plots of $\rho(t) t^{0.5}$
vs $t$ for the \pcpdt with the branching bias near criticality.
The values of $p$ corresponding to three curves are 0.192\;01, 0.192\;02, and 0.192\;03,
respectively, from top to bottom. Since the upper (lower) curve veers up (down) and the middle one
becomes straight, the critical point is estimated as $p_c = 0.192\;02(1)$ with the number
in the parentheses to be the error of the last digit. The same notation is used for numerical
errors throughout the paper.
Inset: Semilogarithmic plot of $\rho(t)/\rho_p(t)$ vs $t$ at criticality
($p=0.192\;02$).
The logarithmic increase at criticality is consistent with the DPCPD
behavior~\cite{PP05a}.}
\end{figure}

Figure~\ref{Fig:DPCPD} summarizes the simulation results using
the system size $2^{23}$. For each curve, 80 independent samples
are collected. Up to the observation time, every sample has
pairs, which minimally
guarantees that the system does not feel the finite size effect.
During simulations, we measured the particle density $\rho(t)$, and
the pair density $\rho_p(t)$.
Figure~\ref{Fig:DPCPD} shows how the density decays near criticality for
the \pcpdt with the biased branching ($\rho\sim t^{-\delta}$ with $\delta=1/2$
in the long-time limit at criticality) and its inset shows the
logarithmic corrections to the power law scaling.
The observed critical behavior is the same as that of the DPCPD~\cite{PP05a},
which reconfirms that the parity conservation is irrelevant in the PCPD.

Now we move to the unbiased \pcpdt with $\sigma_R = \sigma_L = (1-p)/2$ in
Eq.~\eqref{Eq:pcpd}.
In order to study the crossover behavior near the PCPD fixed point,  it is
necessary to measure the critical point of the \pcpdt as accurate as possible.
To this end, we use the effective exponent of the density decay defined as
\begin{equation}
-\delta_\text{eff}(t) \equiv \frac{\ln(\rho(t)) - \ln (\rho(t/m))}{\ln m},
\end{equation}
with $m=10$, which will drift to zero ($-\frac{1}{2}$) in the active (absorbing) phase with time and
at criticality will saturate to the critical decay exponent $-\delta$ of the PCPD
 between $0$ and $-\frac{1}{2}$.
To locate the critical point, we make two assumptions. First, if
$-\delta_\text{eff}(t)$ veers down substantially, we took this as the signal of the
absorbing phase. On the other hand, if the effective exponent veers up and passes
the DP value $\delta_\text{DP}\simeq 0.1595$, we will conclude that the system is in the
active phase. This must be true if $\delta$ for the PCPD is not smaller than $\delta_\text{DP}$
asymptotically, which has been confirmed in all of previous studies for the PCPD.

\begin{figure}[t]
\includegraphics[width=0.45\textwidth]{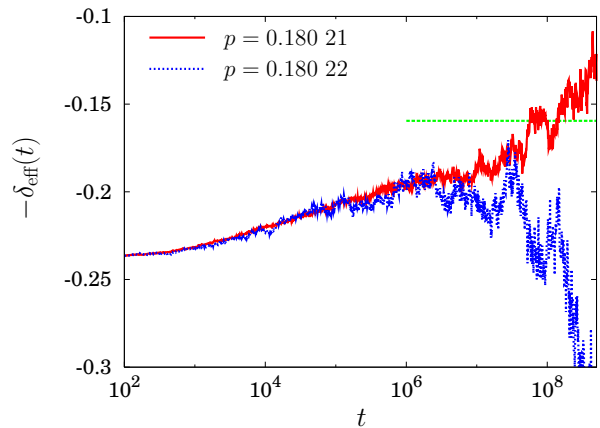}
\caption{\label{Fig:mod2pcpd} (Color online) Plots of the effective exponent
$-\delta_\text{eff}(t)$ vs $t$ for the \pcpdt near criticality
in semi-logarithmic scales. For comparison, the DP value of $-\delta_\text{DP}$
is also shown (straight line segment). Note that the effective exponent for $p=0.180\;21$
passes the DP value around $t=10^8$ and continues to go upwards, while that for $p=0.180\;22$
starts to veer down already around $t=10^7$. Hence the critical point is estimated as
$p=0.180\;215(5)$.}
\end{figure}

We simulated the \pcpdt at $p=0.180~21$ and $p=0.180~22$ with system size
$2^{20}$ up to $t=5 \times 10^8$. The diffusion constant is set as $D=\frac{1}{2}$.
The numbers of independent samples are 120 and 30 for
$p=0.180~21$ and 0.180 22, respectively.
In these simulations, all samples survive up to the observation time in the sense
that there is at least one pair.
The resulting effective exponents are depicted in Fig.~\ref{Fig:mod2pcpd}.
Clearly, the \pcpdt with the annihilation probability $p=0.180~22$ falls
into the absorbing phase. At $p=0.180~21$, the effective exponent
drifts and passes well above the DP value. Hence the critical point should be located between
these two values and  we conclude $p_c  = 0.180~215(5)$ for the PCPD2. Note that
our estimate for $p_c$ is valid, regardless of the true asymptotic exponent value $\delta$ for
the PCPD, unless $\delta$ is much smaller than  $\delta_\text{DP}$.

\section{\label{Sec:models}Crossover models}

In this section, we introduce three models, each of which describes the PCPD-to-DI, the DP-to-DI, and
the PCPD-to-DP crossover, respectively.

\subsection{PCPD-to-DI}

First, introducing the single-particle branching dynamics which preserves the parity
($A \rightarrow 3 A$) to the PCPD2, the crossover from the PCPD to the DI is studied.
With this single-particle dynamics, it is clear that the branching processes by pairs ($2A \rightarrow 4A$)
become irrelevant (a higher-order process) and thus the system must share the critical behavior with the branching
annihilating random walks with two offspring (BAW2)~\cite{TT92}, which belongs to the DI class in one dimension.
We will refer to this model as the PCPD-to-DI crossover model, or simply \Cpcpdt.
To be specific, together with the dynamics in Eq.~\eqref{Eq:pcpd}, the single-particle
branching process is included in a static fashion~\cite{KP95}
\begin{equation}
\emptyset A \emptyset \stackrel{w}{\longrightarrow} AAA.
\label{Eq:baw}
\end{equation}
For a technical reason, the diffusion constant is now set as $D = (1-w)/2$ with
$0\le w \le 1$.

Below is the algorithm how we simulate the \Cpcpdt.
The Monte Carlo simulation begins with a random selection of a particle
among $N_t$ particles at time $t$.
Let us denote the site on which the chosen particle resides by $n$.
With probability $\frac{1}{2}$, we choose one of the direction
(left or right) which will be denoted by $e \in \{1,-1\}$.
If the $(n+e)$-th site is vacant, the particle at site $n$ hops
to that site with probability $1-w$.
With probability $w$,
two offspring are newly placed at two nearest neighbors of
the site $n$, only if the $(n-e)$-th site is also vacant.
On the other hand, if the $(n+e)$-th site is already occupied,
both particles are removed from the system with probability
$p$ or  with probability $1-p$,
we choose two sites in a row, that is, either ($n+2e$, $n+3e$)
or ($n-e$, $n-2e$) with equal probability
and branch two particles there once both sites are empty.
If any of the conditions stated above is not satisfied, nothing happens.
After the above attempt, time increases by $1/N_t$.

\subsection{DP-to-DI}

For comparison, we introduce a crossover model from the DP to the DI to check the difference
of the PCPD and the DP in the crossover behavior to the DI via the same operation $(A\rightarrow 3A)$.
To this end, we need a DP-type model with the parity conservation. One example is the
diffusing \dpt with
$3A\rightarrow 5A$ and $2A\rightarrow 0$. Including $A\rightarrow 3A$, the process $3A\rightarrow 5A$
becomes irrelevant as in the PCPD2 and the model belongs to the DI class. We refer to this model as the DP-DI. The algorithm is identical to the \Cpcpdt except the triplet branching process:
If the $(n+e)$-th site is occupied, then with probability $1-p$ we look up the nearest-neighbor site
of the pair in the right (site $n+(3+e)/2$). If this site is occupied, two particles are generated at nearest neighbors of this triplet, in case both nearest neighbor sites are vacant.

\subsection{PCPD-to-DP via a pair-binding route}

Finally, we consider an alternating diffusion bias in the one dimensional PCPD2.
All particles diffuse with bias in one direction or the other, and neighboring particles
have opposite directional bias. Neighboring particles may be separated by an arbitrary stretch of
empty sites. If a particle hops to the right with rate $(1+w)/2$ and to the left with $(1-w)/2$,
then the hopping rates for neighboring particles are just reversed.
As the particles are created or annihilated pairwise locally in Eq.~\eqref{Eq:pcpd}
and the hardcore repulsion prevents the crossing of particles, each particle can
preserve the bias direction and the alternating diffusion bias in the system is still intact
during the dynamics.  Therefore, one may identify particles with different bias as different species of particles. For example, one may name a particle with diffusion bias to the right as an $A$ particle
and a particle with diffusion bias to the left as a $B$ particle. Then the $AB$ ordering
is preserved if one starts with a configuration with the $AB$ ordering under the
dynamics of $AB\rightarrow\emptyset\emptyset$,
$AB\emptyset\emptyset\rightarrow ABAB$,
$BA\rightarrow\emptyset\emptyset$, and $BA\emptyset\emptyset \rightarrow BABA$
mimicing the PCPD2 dynamics. Without bias ($w=0$), two species
are indistinguishable and we recover the PCPD2.

At $w\neq 0$, this model belongs to the DP class. First, consider the extreme case ($w=1$),
where $A$ ($B$) always jumps to the right (left).
After a transient time, all particles form $AB$ bound pairs which cannot split once bound.
Afterwards, any diffusion is impossible and the dynamics becomes
identical to the contact process (CP) of $X\rightarrow 0$ and $X\rightarrow 2X$ with $X\equiv AB$.
For finite $w$, the biased diffusion will dominate in large scales
and all particles form a little bit loose but still bound $AB$ pairs. Thus we expect the same
critical behavior as in the case $w=1$. Since the CP belongs to the DP class,
so does the model with $w\neq 0$. We emphasize that the fundamental difference between the PCPD and the DP
lies in this pair binding property. The crossover from the PCPD to the DP via
this pair-binding route can be studied near $w\simeq 0$.\

We may map the PCPD2 onto the ferromagnetic Ising spin model by interpreting
a particle as a domain wall of the Ising spins~\cite{KP94,HKPP98,PP08b}. Then an $A$ and $B$ particle
represent a $\uparrow\downarrow$ and $\downarrow\uparrow$ domain wall, respectively.
The alternating bias in diffusion can be interpreted as a symmetry-breaking (SB) field
favoring $\uparrow$ ($\downarrow$) for $w>0$ ($w<0$). So we may call the PCPD2 with finite $w$
as the symmetry-broken \pcpdt (SBPCPD). The crossover
scaling property by the symmetry-breaking (or pair-binding) route may be different
from the conventional one observed in Ref.~\cite{PP06}, similar to various independent routes found in the DI-to-DP crossover~\cite{PP08b}. It is interesting to note that
the conventional crossover is also found by a single-particle
reaction process $(A\rightarrow 2A)$ breaking the parity conservation in the PCPD2.

We remark that the alternating bias (symmetry-breaking field) in the BAW2 also makes two neighboring particles form a bound pair, which triggers the crossover from the DI to the DP. Therefore
the pair-binding process is the common underlying mechanism for the SB routes of the DI-to-DP
and the PCPD-to-DP crossover. However, the corresponding crossover scaling exponents
are different, which depend on the scaling property of the initial fixed points.

\subsection{crossover scaling}

The crossover can be described by the scaling function \cite{LS84,Car96b,PP06},
\begin{equation}
\label{Eq:dpcross}
\rho(t;w,\Delta) = t^{-\delta} {\cal F} (\Delta^{\nu_\|} t,
w^{\mu_\|} t),
\end{equation}
where $\Delta=p-p_c(0)$ with $p_c(0)$ being the critical point at $w=0$ and
$\mu_{||}=\nu_{||}/\phi$ with the crossover exponent $\phi$. The exponents
$\delta$ and $\nu_{||}$ are the density decay and the relaxation time exponent, respectively,
at the initial fixed point ($w=0$).

The crossover scaling function in Eq.~\eqref{Eq:dpcross} implies that
the crossover time scale diverges as $w\rightarrow 0$ such that $\tau_\text{cross}\sim w^{-\mu_\|}$
and the critical amplitude for the density decay at finite $w$'s should scale
as $\rho t^{\delta_\text{T}}\sim w^{\chi}$ with $\chi = \mu_\|(\delta -
\delta_\text{T} )$~\cite{PP06} where $\delta_\text{T}$ is
the density decay exponent at the terminal fixed point ($w\neq 0$).
The crossover exponent $\phi$ can be most accurately measured from studying how the phase boundary $p_c(w)$
approaches the critical point $p_c(0)$ as $w\rightarrow 0$
such that $\Delta_c(w)\sim w^{1/\phi}$ with
$\Delta_c(w)=p_c(w)-p_c(0)$.

\section{\label{Sec:Numerics} Numerical results}

\subsection{\label{sub:PCPD-DI}PCPD-DI}

\begin{figure}[b]
\includegraphics[width=0.45\textwidth]{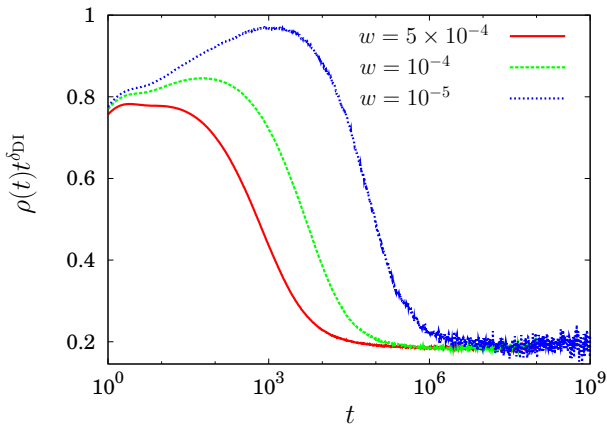}
\caption{\label{Fig:pcpdtodi} (Color online) Plots of $\rho(t) t^{\delta_{\text{DI}}}$ vs
$t$ of the \Cpcpdt for  finite $w$'s at criticality in
semi-logarithmic scales. The smaller $w$ is, the later the system enters the asymptotic scaling regime.
Although $\delta_\text{DI} > \delta_\text{PCPD}$ is evident,
the critical amplitudes do not seem to vary with $w$ up to $w = 10^{-5}$.
}
\end{figure}

This section studies the \Cpcpdt focusing on the crossover exponent $\phi$.
The critical point for the \Cpcpdt with nonzero $w$ can be accurately
measured by using the known DI exponent value $\delta_\text{DI}\simeq 0.285$.
The DI scalings for some $w$'s are shown in Fig.~\ref{Fig:pcpdtodi} and
our findings about the critical points are summarized in Table~\ref{Table:pcpdtodi}.
The crossover time scale seems to diverge as expected in the $w\rightarrow 0$ limit,
which is consistent with the scaling theory prediction of Eq.~\eqref{Eq:dpcross}.
However, the amplitude of the critical decay
($\lim_{t\rightarrow\infty} \rho t^{\delta_\text{DI}}$) does not seem to vary much with $w$.
Similar observation was reported in Ref.~\cite{PP06} for the crossover
from the PCPD to the DP; see Fig.~2 of Ref.~\cite{PP06}.
Of course, this does not mean $\chi = 0$ or
$\delta_\text{PCPD} = \delta_\text{DI}$.
It just implies a rather big correction to scaling which results in
the narrow crossover scaling regime near the
PCPD fixed point~\cite{PP06} .

\begin{figure}[b]
\includegraphics[width=0.45\textwidth]{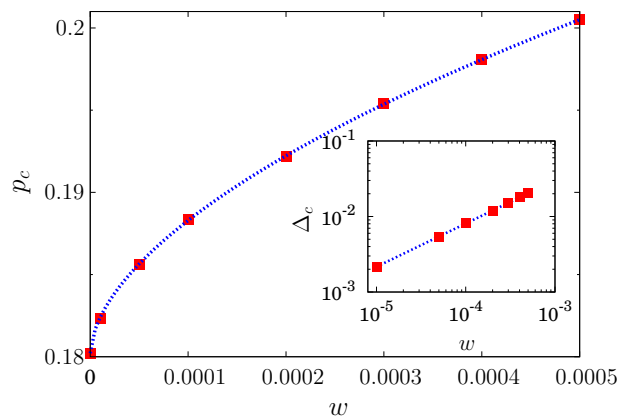}
\caption{\label{Fig:cpb_phase} (Color online) The phase boundary of the \Cpcpdt near
$w=0$. The curves is well fitted by the function
$p_c = p_c(0) + aw^{1/\phi}$ with $1/\phi = 0.57(3)$ and $a\approx 1.61$.
Inset: Plot of $\Delta_c$ vs $w$ in the logarithmic scales.
The slope of the straight line is 0.57.
}
\end{figure}

\begin{table}[t]
\caption{\label{Table:pcpdtodi} The critical points of the \Cpcpdt
for various $w$'s. The numbers in parentheses indicate
the error of the last digits.}
\begin{ruledtabular}
\begin{tabular}{ll}
$w$ & $p_c(w)$\\
\hline
0    & 0.180~215(5)\\
$10^{-5}$  & 0.182~35(5)\\
$5\times 10^{-5}$  & 0.185~6(1)\\
$10^{-4}$  & 0.188~35(5) \\
$2\times 10^{-4}$  & 0.192~2(1)\\
$3\times 10^{-4}$  & 0.195~4(1) \\
$4\times 10^{-4}$  & 0.198~05(5)\\
$5\times 10^{-4}$  & 0.200~5(1)
\end{tabular}
\end{ruledtabular}
\end{table}
Nevertheless, the crossover exponent can be quite accurately estimated from the phase boundary.
Using the data for the critical points in Table~\ref{Table:pcpdtodi}, we estimate the crossover
exponents as $1/\phi = 0.57(3)$ which is summarized in Fig.~\ref{Fig:cpb_phase}.
This value of the  crossover exponent is very close to that reported in Ref.~\cite{PP06}
which studies the crossover from the PCPD to the DP via adding single-particle reaction processes.
We do not think this is a mere coincidence. Rather, we believe that
they are identical.

To argue the equivalence, let us first reexamine the meaning of
the crossover exponent. Consider two universality classes,
say I (initial) and T (terminal), and
an operator triggering the crossover from the I to the T fixed points.
In case of the PCPD-DI,
the I fixed point resides at $(p=p_c(0), w=0)$ and the T fixed point
at $(p=p_c(w^*), w=w^*)$ with $w^*>0$, usually distant from the I fixed point.
There are two relevant operators which make the I fixed point unstable.
One is the crossover operator responsible for the renormalization group (RG) flow
in the direction of $w$ and eventually into the T universality class.
The other is the usual ``thermal'' operator responsible for the RG flow
in the direction of the $p$ axis.

The crossover exponent describes how the RG flow behaves locally near the I fixed point
and is given by the ratio of the scaling dimensions of these two
relevant operators. These scaling dimensions should be decided
entirely in terms of eigenvalues at the I fixed point~\cite{Car96b}.
Hence the T fixed point has no direct role in determining the crossover exponent.
Therefore, whatever the T universality class is, the crossover exponent
at the I fixed point may not change if
the scaling dimensions of two relevant operators remain the same.

Now let us move on to our problem.
We should emphasize that the PCPD is insensitive to the existence
of the parity conservation. Therefore, the PCPD fixed point does not feel
any intrinsic difference when we add one of the single-particle reaction
processes of $A\rightarrow nA$ for any $n$.
If the PCPD can be described by a well-defined fixed point, then
the scaling dimension of the crossover operator associated with the $A\rightarrow nA$ process
will not depend on $n$. Of course, the terminal (T) fixed point is sensitive to the
parity conservation and thus belongs to the DP for odd $n$ or to the DI for even $n$.\
This distinction cannot be achieved by a perturbative RG method (like the crossover exponent study),
but by studying the global RG flows.

Hence, we conclude that the crossover exponent should be the same for both
the PCPD-to-DP and the PCPD-to-DI crossover via a single-particle reaction route.
Our numerical confirmation of this equivalence, in turn, provides another evidence
that the PCPD has a well-defined fixed point distinct from the DP.

\subsection{DP-DI}

\begin{figure}[t]
\includegraphics[width=0.45\textwidth]{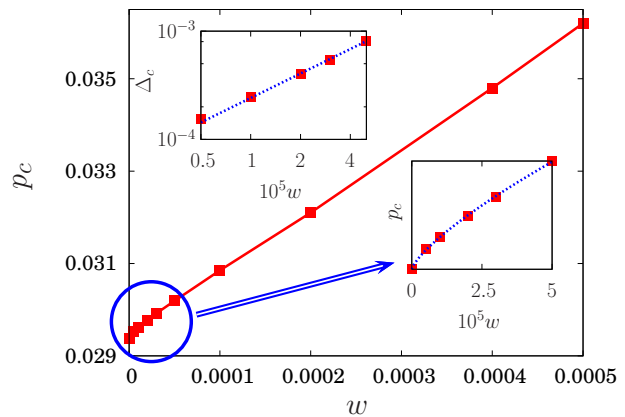}
\caption{\label{Fig:ctb_phase} (Color online) The phase boundary of the \Cdpt near
$w=0$. For $w \ge 10^{-4}$, the phase boundary looks almost a straight line,
but the close-up of the region of $w \le 5 \times 10^{-5}$ (panel
located at the right bottom) shows a curvature which is fitted with
the crossover exponent $1/\phi = 0.73(4)$.  The panel at the left top is the
plot of $\Delta_c$ vs $10^5 w$ in the logarithmic scales.
The slope of the straight line is 0.75.
}
\end{figure}

For comparison, we study the DP-DI model to study the
crossover behavior from the DP to the DI
mediated by the same operation ($A\rightarrow 3A$).
We again confirmed that this hybid-type multiple reaction model (the \Cdpt at $w=0$) does belong to the DP class (not shown here).
As before, the critical points for finite $w$'s are determined, exploiting the
DI critical scaling.
The results are summarized in Table~\ref{Table:dptodi} and in Fig.~\ref{Fig:ctb_phase}.

\begin{table}[t]
\caption{\label{Table:dptodi} The critical points of the \Cdpt
for various $w$'s.}
\begin{ruledtabular}
\begin{tabular}{ll}
$w$ & $p_c \text{ of the \Cdpt}$\\
\hline
0    &  0.029~376(1)\\
$5\times 10^{-6}$&0.029 530(5)\\
$10^{-5}$  & 0.029 620(5) \\
$2\times 10^{-5}$&0.029 775(5)\\
$3\times 10^{-5}$&0.029 920(5)\\
$5\times 10^{-5}$&0.030 185(5)  \\
$10^{-4}$    &0.030 85(5)\\
$2\times 10^{-4}$  &0.032~1(1) \\
$4\times 10^{-4}$  & 0.034~8(1)\\
$5\times 10^{-4}$  & 0.036~20(5)
\end{tabular}
\end{ruledtabular}
\end{table}
\begin{figure}[b]
\includegraphics[width=0.45\textwidth]{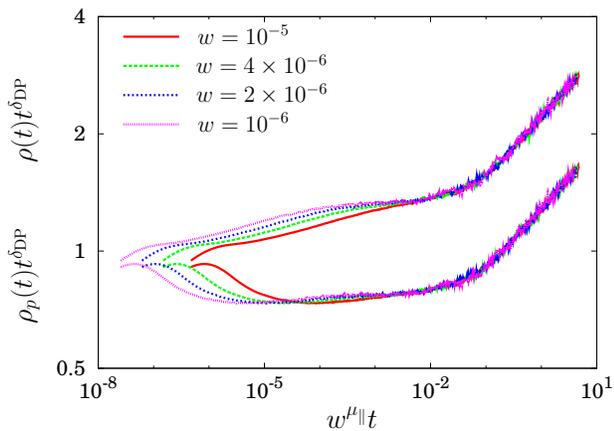}
\caption{\label{Fig:dpcol} (Color online) Scaling collapse for the particle ($\rho$)
and pair densities ($\rho_p$) in log-log scale with $\Delta=0$; see Eq.~\eqref{Eq:dpcol}.
Both particle and pair densities
are collapsed into a respective single curve in the asymptotic regime.
}
\end{figure}

If one stops the analysis at $w= 10^{-4}$,  one might conclude
that the phase boundary meets the vertical axis with a finite angle, and therefore
the crossover exponent is $1$. However, our more elaborated study
 with sufficiently small $w$'s shows that the phase boundary
has a singularity with the crossover exponent $1/\phi = 0.73(4)$;
see Fig.~\ref{Fig:ctb_phase}.

Now turn back to the scaling function in Eq.~\eqref{Eq:dpcross}.
If we fix $\Delta  = 0$, the crossover scaling function
takes the form
\begin{equation}
\rho(t) t^{\delta_\text{DP}} = {\cal C} (w^{\mu_\|} t),
\label{Eq:dpcol}
\end{equation}
where the numerical value of $\mu_\| = \nu_\|^\text{DP}/\phi = 1.26(7)$
with $\delta_\text{DP}\simeq 0.16$ and $\nu_\|^\text{DP}\simeq 1.73$.
In Fig.~\ref{Fig:dpcol}, we depicted the scaling collapse for the
particle and pair densities for $10^{-6} \le w \le 10^{-5}$.
All curves are collapsed into a single curve in the asymptotic regime,
which means that the scaling function in Eq.~\eqref{Eq:dpcol}
describes the behavior near $w=0$ properly with the
crossover exponent obtained before.
Hence we conclude that the crossover exponent from  the DP
to the DI is $1/\phi = 0.73(4)$ or $\phi = 1.37(8)$ which
is different from that obtained from the \Cpcpdt.

\subsection{\label{sub:sbpcpd} SBPCPD}

Finally, we study the SBPCPD where the alternating diffusion bias is included in the PCPD2.
We expect the crossover from the PCPD to the DP via a pair-binding route, which
should be independent from that via a single-particle reaction route.

The strategy to find the critical points for finite $w$'s is the same as before, i.e.,
exploiting the power law decay of the density with the
DP critical exponent $\delta_\text{DP}$ at criticality.
The results are summarized in Table~\ref{Table:sbpcpd}.
As can be seen from the table, the critical points for finite $w$'s are
quite close to the critical point of the PCPD2. Hence, the accurate
estimation of the critical points are necessary to estimate the
crossover exponent with a reasonable precision. To this end, we simulated
large systems ($N=2^{20}$) for long times (up to $t=2\times 10^7$ for
$w=10^{-2}$ and $t= 5 \times 10^8$ for $w=10^{-3}$).

\begin{figure}[t]
\includegraphics[width=0.45\textwidth]{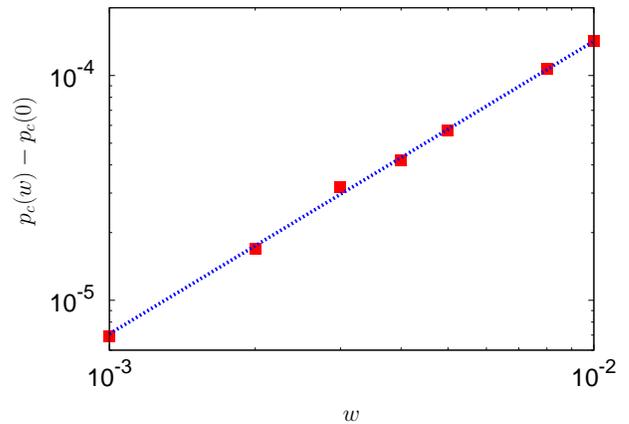}
\caption{\label{Fig:pbSB} (Color online) $p_c(w) - p_c(0)$ vs $w$ plot of
the SBPCPD in log-log scales. The symbols are results from simulation and the straight
line is result of the fitting using the trial function $p_c(w)  = p_c(0) + a\times
w^{1/\phi}$ with three fitting parameters $p_c(0)$, $a$, and $1/\phi$.
The fitting results are $p_c(0) \approx 0.180~214~7$ and $1/\phi \approx 1.23$.
}
\end{figure}

As before, we measure the crossover exponent from the singular behavior of
the phase boundary. Rather than using the critical point of the PCPD2 to
calculate $\phi$, we fit the phase boundary using the trial function
$p_c(w) = p_c(0) + a\times w^{1/\phi}$
with three fitting parameters $p_c(0)$, $a$, and $1/\phi$. The fitting using
the whole data in Table~\ref{Table:sbpcpd} except $p_c(0)$, gives
$p_c(0) \approx 0.180~214~7$ and $1/\phi = 1.23$.
These results are summarized
in Fig.~\ref{Fig:pbSB}.
If we fix $p_c(0)$ between 0.182~21 and 0.182~22 and use $a$ and $1/\phi$
as two fitting parameters, the crossover exponent
varies from $1/\phi=1.13$ to $1.33$. So we estimate the crossover exponent
as $1/\phi = 1.23(10)$.
\begin{table}[t]
\caption{\label{Table:sbpcpd} The critical points of the SBPCPD.}
\begin{ruledtabular}
\begin{tabular}{ll}
$w$ & $p_c(w)$\\
\hline
0.     &0.180~215 (5) \\
$10^{-3}$  &0.180~222~5 (25) \\
$2\times 10^{-3}$  &0.180~235 (5) \\
$3\times 10^{-3}$  &0.180~250 (5) \\
$4\times 10^{-3}$  &0.180~260 (5) \\
$5\times 10^{-3}$  &0.180~275 (5) \\
$8\times 10^{-3}$  &0.180~325 (5) \\
$10^{-2}$  &0.180~360 (10)
\end{tabular}
\end{ruledtabular}
\end{table}

\begin{figure}[b]
\includegraphics[width=0.45\textwidth]{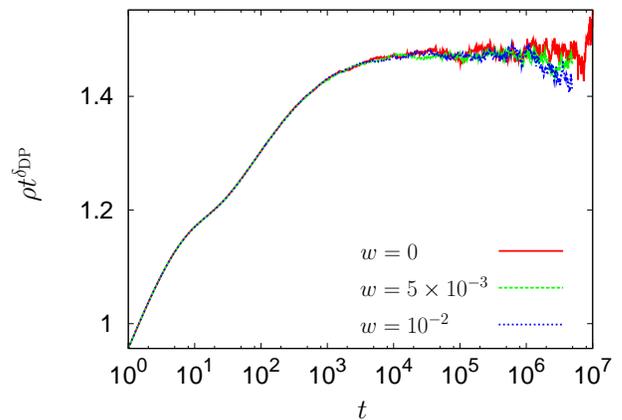}
\caption{\label{Fig:sbdpcol} (Color online) Plots of $\rho (t) t^{\delta_\text{DP}}$ vs
$t$ near criticality for the ``crossover'' model from the DP to the DP mediated by the
symmetry breaking field for $w=0$ ($p=0.031~405$), $5\times 10^{-3}$
($p=0.031~409$), and $10^{-2}$ ($p=0.031~419$).
No clear distinction among curves is observed, which is the characteristics of the
``crossover'' model within the same universality class.}
\end{figure}

For comparison, we consider the {\tt qp22} ($4A \rightarrow 6A$, $2A\rightarrow 0$),
belonging to the DP class. Adding the alternating diffusion bias $w$ (symmetry-breaking field
in the dual spin language), all particles form bound pairs ($X\equiv 2A$), of which
the dynamics is given by $2X\rightarrow 3X$ and $X\rightarrow 0$. Therefore, we expect
the DP scaling even at nonzero $w$. There is no true crossover between the same (DP)
universality classes and so there are no diverging time scale ($\mu_\|=0$) and
also no singular behavior of the critical amplitude ($\chi=0$) as $w\rightarrow 0$.
Figure~\ref{Fig:sbdpcol} confirms our prediction.

\section{\label{Sec:sum}Summary}

To summarize, we introduced and studied two types of crossover models from the
pair contact process with diffusion with the parity conservation (PCPD2).
We study the crossover from the PCPD  by adding
the parity-conserving single-particle process to the directed Ising (DI) class and
by adding the alternating bias (symmetry-breaking field in the spin language)
to the directed percolation (DP) class. For comparison, we also explored the DP-to-DI
crossover by studying the diffusing {\tt tp22} with the parity-conserving single-particle process.

The crossover exponents were found numerically, which
read $1/\phi_1 = 0.57(3)$, $1/\phi_2=0.73(4)$, and $1/\phi_3 = 1.23(10)$
respectively
for the PCPD-to-DI, the DP-to-DI, and the PCPD-to-DP with the alternating bias.
First, we argue that $\phi_1$ is identical to the PCPD-to-DP via a single-particle reaction route,
which is numerically confirmed. The new crossover exponent $\phi_2$ found for
the DP-to-DI crossover is clearly distinct from $\phi_1$, which supports the non-DP nature
of the PCPD fixed point. Third, we found another independent crossover exponent $\phi_3$
for the PCPD-to-DP crossover, which is ascribed to the pair-binding route different
from the conventional single-particle reaction route. Nontriviality of $\phi_3$ as well as $\phi_1$
provides another evidence to support that the PCPD is distinct from the DP.

\acknowledgments
SCP would like to acknowledge the support by DFG within SFB 680 {\it Molecular Basis of Evolutionary Innovations} and by the Korea Institute for Advanced Study
(KIAS) where the idea was formed and a part of the work was
completed. Most of computation was carried out using KIAS supercomputers.


\begin{thebibliography}{99}
\bibitem{com1} Lowest elementary excitations for reactions are pairs for the PCPD-type models. Similar analogy can be considered for general multiple reaction models.
\end{thebibliography}
\end{document}